\documentclass[runningheads,fleqn]{svmult}
\usepackage{makeidx}   % allows index generation
\usepackage{graphicx}  % standard LaTeX graphics tool
\usepackage{subeqnar}  % subnumbers individual equations
\usepackage{multicol}  % used for the two-column index
\usepackage{taphys}    % flushleft layout of captions etc.
\makeindex             % used for the subject index
\newcommand{\beq}{\begin{equation}}
\newcommand{\eeq}{\end{equation}}
\newcommand{\bk}{\mathbf{k}}
\newcommand{\pdag}{{\phantom{\dagger}}}
\begin{document}
\title*{Numerical Renormalization Group Analysis of Interacting Quantum Dots}
\toctitle{Numerical Renormalization Group Analysis of Interacting Quantum Dots}
% allows explicit linebreak for the table of content
%
%
\titlerunning{Numerical Renormalization Group}
% allows abbreviation of title, if the full title is too long
% to fit in the running head
%
\author{Walter Hofstetter}
\authorrunning{Walter Hofstetter}
% if there are more than two authors,
% please abbreviate author list for running head
%
%
\institute{Theoretische Physik III, Elektronische Korrelationen und Magnetismus, 
Universit\"at Augsburg, 86135 Augsburg}

\maketitle              % typesets the title of the contribution

\begin{abstract}
Wilson's Numerical Renormalization Group (NRG) is so far the only nonperturbative 
technique that can reliably access low--energy properties of quantum impurity systems. 
We present a recent  extension of the method, the \mbox{DM--NRG}, which yields  
highly accurate results for dynamical quantities at arbitrary frequencies 
and temperatures. 
As an application, we determine the spectrum of a quantum dot in an external magnetic field. 
Furthermore, we discuss magnetic impurities with orbital degeneracy, 
which have been inferred in 
recent experiments on quantum dots in an Aharonov-Bohm geometry. 
It is demonstrated that for spinless electrons,   
interference between neighbouring levels sets the low--energy scale 
of the system. Switching on an external field leads to a remarkable 
crossover into a regime dominated by \emph{orbital} Kondo screening. 
We predict that the broadening--induced level splitting should 
be clearly visible in measurements of the optical absorption power. 
A more general model including the electron spin is studied within an 
extended two-band NRG procedure. We observe competition between interference 
and Kondo screening, similar to the situation in two-impurity models (RKKY).

\end{abstract}

\section{Introduction}
Quantum impurity models and their low-temperature properties are 
of central importance in condensed matter physics. 
They show characteristic many-body effects like the screening 
of a local moment by conduction electrons (the Kondo effect) 
which was first observed in measurements on dilute magnetic impurities 
in metals (see \cite{Hewson93}).
More recently, artificial nanostructures (quantum dots \cite{kondo-exp} 
or surface atoms probed by STM \cite{Li98,Manoharan00}) 
with tunable parameters provided new representations
of the Anderson or Kondo model \cite{Anderson61,Kondo64}. 
In theory, a very fruitful line of research was opened by the development 
of dynamical mean-field theory (DMFT) \cite{Metzner89} where correlated lattice systems 
are mapped onto effective impurity models which are then accessible 
in a controlled way \cite{Georges96}.

In this article, we focus on semiconductor nanostructures, where 
-- at the moment -- experiments with the highest level of control 
can be performed. 
Electronic transport through ultra-small quantum dots, where the charging
energy is the largest energy scale, has been studied extensively over 
the last few years \cite{curacao}. Due to the quantization of charge the
transport is dominated by Coulomb blockade. 
More recently, experiments revealed that the Kondo effect leads to 
an enhancement of the conductance -- the  \emph{zero bias anomaly} -- 
in the Coulomb blockade regime \cite{kondo-exp,Simmel99} 
as predicted some time ago \cite{kondo-theo}.

Theoretical modelling of these systems is usually based on the Kondo or Anderson
Hamiltonians \cite{Anderson61,Kondo64} describing a localized spin (orbital)
which is coupled to one or several conduction electron reservoirs. 
In the regime of interest, this coupling is the usually the smallest energy scale.  
It was realized very early \cite{Kondo64} that a treatment of these models
based on perturbation theory fails due to logarithmic divergences below a characteristic 
temperature scale, the \emph{Kondo temperature} $T_K$. 

Solution of the Kondo problem at $T \ll T_K$ thus required a \emph{non--perturbative} technique, 
which was provided by Wilson's pathbreaking \emph{Numerical Renormalization Group} (NRG) 
\cite{Wilson75}. 
It proved to be very successful in clarifying the low--energy properties of various 
impurity problems \cite{Krishnamurthy80,Sakai89,Costi94,Bulla98,Hofstetter00b}, 
and it will be the method of choice for analyzing more complex 
quantum dot systems. 

In this article we give a short introduction to the NRG technique, including in particular 
a recently developed density--matrix formalism (the DM--NRG \cite{Hofstetter00a}) 
necessary for a reliable calculation of dynamical quantities. 
Using the new algorithm, impurity spectra are calculated for a quantum dot in an external magnetic field. 
Finally, we discuss the subtle interplay between interference and interaction 
which arises in a quantum dot with orbital degeneracy.

\section{Generalized Numerical Renormalization Group}
In the following, we consider the Anderson Hamiltonian \cite{Anderson61}
\beq   \label{Anderson_Hamiltonian}
H_{\rm and} = \sum_{\bk \sigma} \epsilon_{\bk}\, c^\dagger_{\bk \sigma}\, 
c^\pdag_{\bk \sigma} + \sum_{\bk \sigma} V_{\bk}\, \left(f^\dagger_\sigma\, 
c^\pdag_{\bk \sigma} + h.c. \right) + U\, n_{f \uparrow}\, 
n_{f \downarrow}  + \epsilon_f\, n_f - h\,S_f^z 
\eeq
where the \emph{hybridization} $\Gamma(\omega) = 2\pi \sum_k |V_k|^2\, \delta(\omega - \epsilon_k)$ 
between the $f$ impurity and the reservoir electrons 
$c_{\bk \sigma}$ is balanced by a local Coulomb repulsion $U$ which suppresses double occupancy 
of the impurity. In addition, a local magnetic field $h$ is coupled to the impurity spin $S^z_f$. 
Units are chosen as $\hbar = k_B = \mu_B = g = 1$ and the half bandwidth 
is given by $D=1$. 

\begin{figure}
\begin{center}
\includegraphics[width=.6\textwidth]{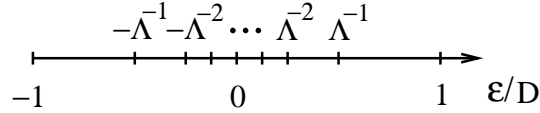}
\end{center}
\caption[]{Logarithmic discretization of the conduction band.}
\label{fig:log_disc}
\end{figure}

\begin{figure}
\begin{center}
\includegraphics[width=.6\textwidth]{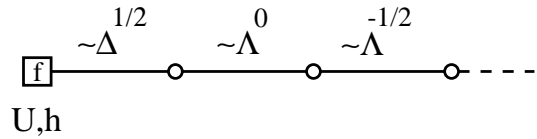}
\end{center}
\caption[]{Linear chain representation of the Anderson impurity Hamiltonian.}
\label{fig:linear_chain}
\end{figure}

\begin{figure}
\begin{center}
\includegraphics[width=.5\textwidth]{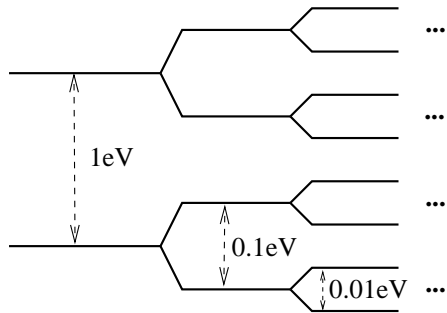}
\end{center}
\caption[]{Iterative diagonalization of the impurity problem.}
\label{fig:iterative_diag}
\end{figure}

The key idea introduced by Wilson is the \emph{logarithmic discretization} of the 
conduction band shown in Fig.~\ref{fig:log_disc}, where each energy scale is represented 
by a single fermionic degree of freedom. After performing a Lanczos transformation 
(for details see \cite{Wilson75,Hofstetter00c})  
the conduction band can be written as a linear chain shown in Fig.~\ref{fig:linear_chain}
\beq
H_{\rm and} = \sum_{n=0 \atop \sigma}^{\infty} \epsilon_n\, 
\left(d^\dagger_{n \sigma}\, d^\pdag_{(n+1) \sigma} + h.c. \right) 
\eeq
with hopping coefficients decaying \emph{exponentially} as $\epsilon_n \sim \Lambda^{-n/2}$.
In this representation, the impurity is only coupled to the 
\emph{maximally localized} reservoir state 
\mbox{$d_0 = (1/\sqrt{N}) \sum_k c_k$}.
The transformed model -- while still a nontrivial many--body problem -- can now be solved by 
iterative diagonalization, keeping in each step only the lowest, most relevant levels. 
This procedure resembles the one employed in calculating atomic spectra 
and is illustrated in Fig.~\ref{fig:iterative_diag}. 
Additional symmetries like the conservation of the total charge and components of the 
total spin can be invoked in order to simplify the remaining matrix algebra. 
In the single--band case considered here, matrix size is not a problem, while 
for calculations with two (different) reservoirs the use of symmetries can be vital 
in order to render the problem manageable. 

In the iterative diagonalization scheme,  
the number of iterations corresponds to the temperature one 
is interested in according to $T_N = c \,\Lambda^{-(N-1)/2} $,  where $c$ is a constant 
of order one. For calculating static thermodynamic expectation values,  
all necessary
information is thus obtained because only excitations on the scale 
$T_N$ are relevant. As an example, consider the impurity magnetization 
\beq
<S^z_f>_T = Z^{-1}\sum_m e^{-\beta E^N_m}\, <m| S^z_f |m>_N
\eeq
where the $|n>$ are the many-particle eigenstates of $H$ and $Z$ is the partition function.
Due to the Boltzmann factor, higher excitations -- already lost in iteration $N$ -- 
can be safely neglected at this point. 

The situation changes completely when we consider a \emph{dynamical} quantity like 
the spin--resolved spectral density 
\beq    \label{rho}
A_{\sigma}(\omega) = 
\sum_{n m} |<m| f^\dagger_{\sigma} |n>|^2 \, 
\delta\left(\omega - E_{m} + E_{n}\right)\, 
\frac{e^{-\beta E_m} + e^{-\beta E_n}}{Z}.
\eeq
Obviously, spectral information at frequencies $\omega \gg T_N$ 
requires matrix elements between low-lying states and excitations
which in iteration $N$ are not available anymore (they have already been lost by truncation). 
In order to deal with this situation, the following \emph{two--stage} procedure 
has to be employed: 

(1) NRG iterations are performed down to the temperature $T_N$ of
interest, in particular we choose $T_N \ll T_K$ to calculate 
ground-state properties. 
In each iteration step, we keep the information on the transformation 
between one set of eigenstates and the next, i.e.~we save the
corresponding unitary matrix. After obtaining the relevant excitations 
at temperature $T_N$ one can define the density matrix 
\beq
\hat{\rho} = Z^{-1} \sum_m e^{-E^N_m/T_N}\, |m>_{N} <m|
\eeq
which completely describes the physical state of the system. 
The equilibrium Green's function can be written as
\beq   \label{new_Green}
G_{\uparrow}(t) = i \theta(t) {\rm Tr} \left(\hat{\rho}\, 
\left\{f^\pdag_{\uparrow}(t), f^\dagger_{\uparrow}(0)\right\}\right)
\eeq
\begin{figure}
\begin{center}
\includegraphics[width=.6\textwidth]{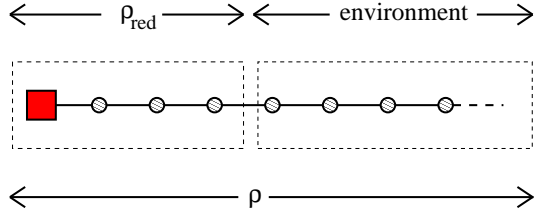}
\end{center}
\caption[]{Reduced density matrix obtained by tracing out ``environment'' 
degrees of freedom of the chain.}
\label{fig:reduced_DM}
\end{figure}

(2) Now we repeat the iterative diagonalization for the same
parameters. Each iteration step $N'$ yields the single-particle 
excitations (and matrix elements of $f^\dagger$) 
relevant at a frequency $\omega \sim T_{N'}$. 
But instead of using (\ref{rho}), we now 
employ (\ref{new_Green}) and evaluate the spectral
function with respect to the correct \emph{reduced density matrix} \cite{Feynman72}:
As depicted in Fig.~\ref{fig:reduced_DM}, the complete chain is split 
into a smaller cluster of length $N'$ and an \emph{environment}
containing the remaining degrees of freedom. In the product basis 
of these two subsystems, the full density matrix has the form
\beq
\hat{\rho} = \sum_{m_1 m_2 \atop n_1 n_2} 
\rho_{m_1 n_1, m_2 n_2} |m_1>_{\rm env} |n_1>_{\rm sys} 
<n_2| <m_2|
\eeq
which is in general not diagonal anymore.
Performing a partial trace on the environment then yields the 
density submatrix 
\beq
\hat{\rho}^{\rm red} = \sum_{n_1 n_2} \rho^{\rm red}_{n_1 n_2} 
|n_1>_{\rm sys} <n_2| \ \ \ \ , \ \ \ \ 
\rho^{\rm red}_{n_1 n_2} = \sum_m \rho_{m n_1, m n_2}
\eeq
This projection is easily done using the previously stored 
unitary transformation matrices. 
Note that $\rho^{\rm red}$ -- defined only on the shorter chain -- 
contains all the relevant information about the 
quantum mechanical state of the \emph{full} system.

\begin{figure}
\begin{minipage}{0.5\linewidth}
\includegraphics[width=0.95\textwidth]{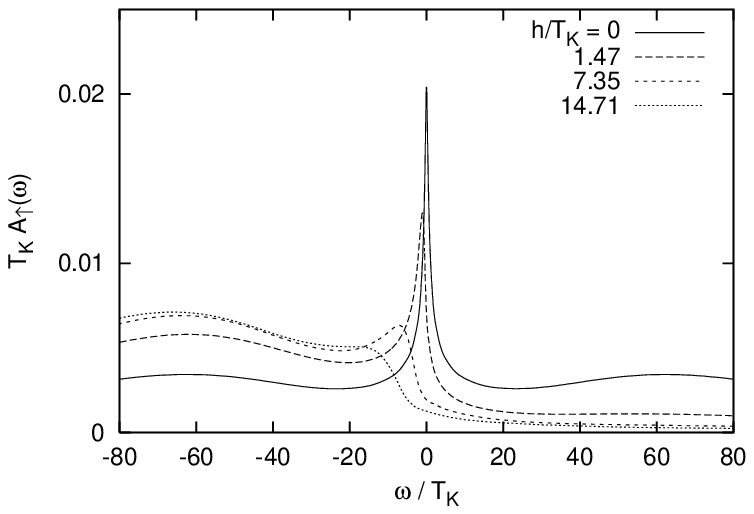}
\end{minipage}
\begin{minipage}{0.5\linewidth}
\includegraphics[width=0.95\textwidth]{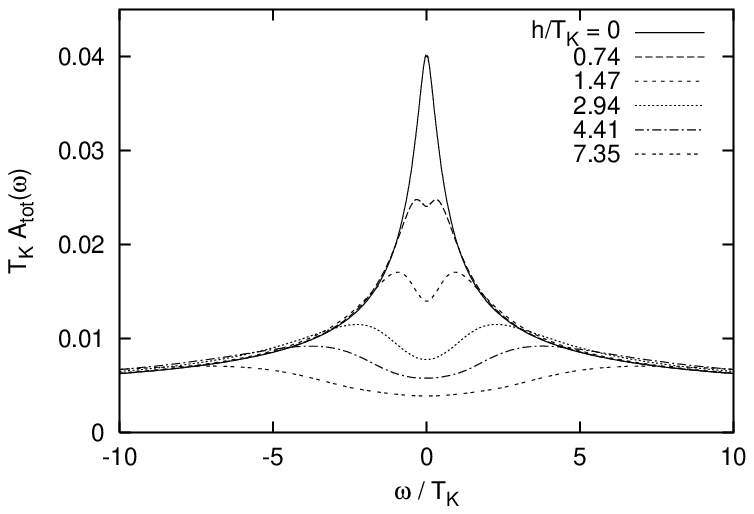}
\end{minipage}
\caption[]{Spin--dependent (left) and total (right) 
impurity spectral density at zero temperature for $\Gamma=0.02$, $U=5\Gamma$, 
and $\epsilon_f = -2.5\Gamma$. The Kondo temperature is $T_K = 6.8\times 10^{-4}$.
}
\label{fig:spectrum}
\end{figure}

The single--particle spectrum calculated in this way is shown in Fig.~\ref{fig:spectrum}. 
With increasing magnetic field, the Kondo resonance 
is suppressed and eventually merges with the lower atomic level. 
Regarding the total density of states (DOS)  
$A(\omega) = \sum_{\sigma} A_{\sigma}(\omega)$, the Kondo peak 
is split by the field and the DOS at the Fermi level
strongly reduced. This effect has been observed directly 
in measurements of the differential conductance through a quantum dot 
\cite{kondo-exp}.

\section{Interference and Interaction in Multi--Level Dots}
After the discussion of the spin--degenerate Anderson impurity model 
(\ref{Anderson_Hamiltonian}) in the last section, we now consider 
the effect of \emph{orbital} degeneracy. We will first study 
a dot consisting of two levels without spin or, 
equivalently, two dots in an Aharonov-Bohm (AB) geometry with one
level per dot in the presence of an interdot Coulomb repulsion $U$. 
Such a system is of fundamental interest since the
two possible paths through the dot (via level $1$ or
$2$) can interfere with each other. The interference can be controlled
by an AB flux and has attracted much interest due to
the possibility of realizing AB interferometers \cite{Yacoby97}
or using the coherent properties in connection with quantum
computing \cite{loss}.
Furthermore, there is enhanced experimental
interest to study quantum dots in the strong tunneling regime where 
the level broadening is of the order of the level spacing. In this
case, transport is inevitably controlled by multi-level physics.
\begin{figure}
\begin{center}
\includegraphics[width=.6\textwidth]{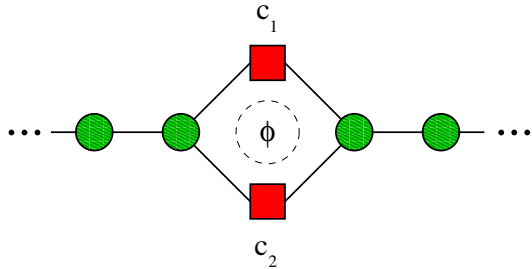}
\end{center}
\caption[]{Two quantum dots in an Aharonov--Bohm geometry.}
\label{fig:AB}
\end{figure}

Let us first discuss the case of \emph{spinless} electrons, assuming 
e.g. a large Zeeman splitting \cite{Boese00}. 
The Hamiltonian is then written as  
\beq
H =  \sum_{kr}  \varepsilon_{kr} a^\dagger_{kr} a_{kr}  + 
\sum_{rkj} (V^r_j a^\dagger_{kr} c_j + {\mathrm h.c.}) + 
\sum_j \varepsilon_j c^\dagger_j c_j + U n_1 n_2
\label{AB_Hamiltonian}
\eeq
where $j=1,2$ labels the two levels and the dot is connected to two reservoirs $r=L,R$ 
via tunnel barriers. 
Note that the index labelling the dots is not present in the reservoirs -- 
this model contains no conserved quantum number corresponding to spin, 
unlike previous studies \cite{Inoshita93,Pohjola97,Izumida97,Izumida98}. 
The tunnel matrix elements are assumed to be real except for an
AB-phase, i.e. we attach a phase factor $e^{i\phi}$ to
$V_2^L$. The level broadening is defined by $\Gamma_j^r = 2\pi |V_j^r|^2\rho_0$, 
where $\rho_0$ is the density of states in the leads, which we assume to be constant 
in the energy range of interest. 
The total broadening of each level is therefore given by 
$\Gamma = \Gamma^L_j + \Gamma^R_j$. 

\begin{figure}
\begin{minipage}{0.5\linewidth}
\begin{center}
\includegraphics[width=0.95\textwidth]{autfig8.eps}
\end{center}
\end{minipage}
\begin{minipage}{0.5\linewidth}
\begin{center}
\includegraphics[width=.8\textwidth, angle=-90]{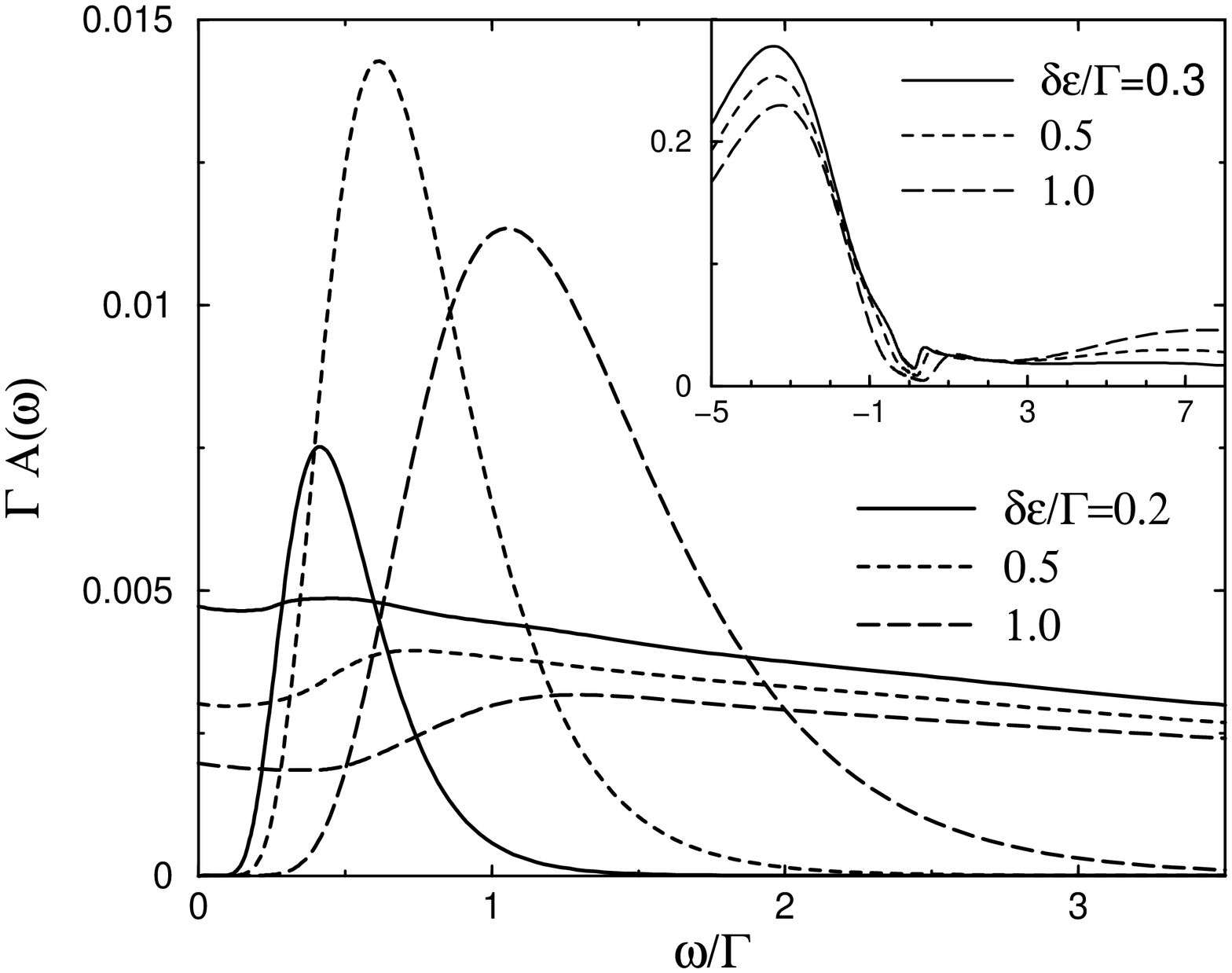}
\end{center}
\end{minipage}
\caption[]{Left: tunnel splitting visible in the total spectral function for parameters 
$U=16\Gamma$, $\epsilon_1 = \Phi = T = 0$ and different positions of the 
second levels $\epsilon_2$. Right: 
Absorption power (broken lines) vs. single--particle spectral density (full lines) for 
$U=50 \Gamma$, $\epsilon_1 = -10\Gamma$ and $\Phi = T = 0$. In the inset, 
$U = 10 \Gamma$ and $\epsilon_1 = -3.5 \Gamma$ \cite{Boese00}.}
\label{fig:absorption}
\end{figure}

Since both levels overlap with the reservoirs, they have an effective 
overlap matrix element $\Delta$, which induces a level splitting 
$\delta\tilde{\epsilon} = \sqrt{\delta\epsilon^2 + |\Delta|^2}$
where $\delta\epsilon = \epsilon_2 - \epsilon_1$ denotes the bare level spacing.
Within second--order perturbation theory it is established that 
$\Delta$ vanishes for the noninteracting dot, while 
\beq
\Delta\sim {\sqrt{\Gamma_1^R\Gamma_2^R}+
            \sqrt{\Gamma_1^L\Gamma_2^L}e^{i\phi}\over \pi}
            \ln{(U/\omega_c)}
\eeq
in the case of strong on--site repulsion $U \gg |\epsilon|,\Gamma$. 
For positive level energies, the tunnel splitting can be observed directly 
as a shift of the upper level position (see Fig.~\ref{fig:absorption}) in 
the total spectral density 
\beq
A(\omega) = -\frac{1}{\pi}\, \sum_{i,j=1,2} {\rm Im} G_{ij}(\omega^+).
\eeq
For low lying levels $\epsilon < -\Gamma$ and a large Coulomb repulsion,  
the dot is singly occupied at low temperatures. 
In this case, the effective level splitting shows up in a many-body resonance (``shoulder'') in
the spectral density at a \emph{positive} frequency  
$\omega \sim \delta\tilde{\epsilon}$, see Fig.~\ref{fig:absorption}. 
This new many--body energy scale can be seen most clearly in a microwave absorption experiment, 
where transitions between the two dot levels are induced due to the dipole operator 
\mbox{$\hat{O} = c^\dagger_1 c^\pdag_2 + c^\dagger_2 c^\pdag_1$}. 
The corresponding spectral density 
\beq
\rho_{\rm abs}(\omega) = Z^{-1} \sum_{mn} |<n|\hat{O}|m>|^2\, 
\delta\left(\omega + E_n - E_m\right)\, \left(e^{-\beta E_n} - e^{-\beta E_m}\right)
\eeq
displays a well--pronounced resonance at the frequency $\delta\tilde{\epsilon}$ 
(see Fig.~\ref{fig:absorption}).

\begin{figure}
\begin{minipage}{0.5\linewidth}
\begin{center}
\includegraphics[width=.85\textwidth, angle=-90]{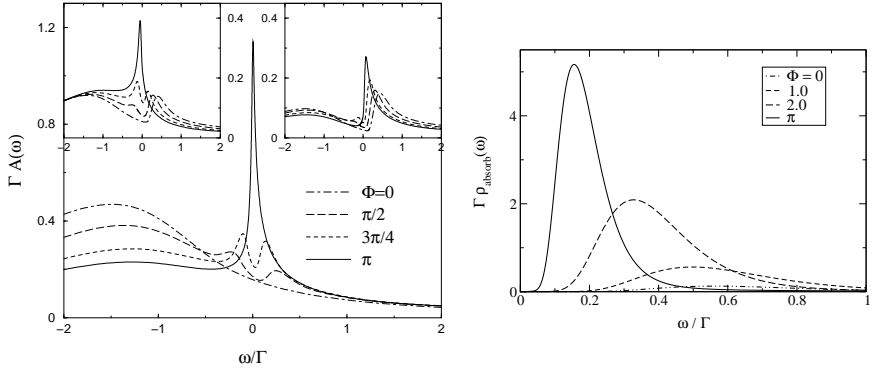}
\end{center}
\end{minipage}
\begin{minipage}{0.5\linewidth}
\begin{center}
\includegraphics[width=.9\textwidth]{autfig11.eps}
\end{center}
\end{minipage}
\caption[]{Left: Effect of the AB--phase $\phi$ on the single--particle spectrum. 
The total spectral density is shown for $\epsilon_1 = -1.6\Gamma$, 
$\delta_\epsilon = 0$, $U = 8.1\Gamma$, and $T=0$. 
Insets: Partial spectral densities for levels $c_1$ (left) and $c_2$ (right). 
Same parameters as above, but with a finite level splitting 
$\delta\epsilon = 0.08 \Gamma$. 
Right: Absorption power for $U=33\Gamma$, $\epsilon_1 = -3.3\Gamma$, 
$\delta\epsilon = 0.16\Gamma$, $T=0$ and different values of $\phi$.}
\label{fig:Kondo}
\end{figure}

For $\phi=\pi$ and $\Gamma^R_j=\Gamma^L_j$, the tunnel splitting is 
zero, and the system is shown to be equivalent to an Anderson model 
with Zeeman splitting $\delta\epsilon$. 
This can be seen most easily by introducing the new levels 
\begin{equation}
c_{\rm even(odd)} = (1/\sqrt{2}) (c_1 \pm c_2)
\end{equation}
which are then coupled to the 
right and left reservoirs, respectively.
In this way, an \emph{orbital} Kondo effect can be realized in a quantum dot even in the absence 
of an \emph{a priori} conserved quantum number like spin. 
The crossover between the interference-- and Kondo--dominated regimes upon increase 
of the AB--phase $\phi$ is most clearly seen in the single particle spectrum 
(Fig.~\ref{fig:Kondo}) 
where the Kondo resonance gradually develops as $\phi\to\pi$. 
Experimentally, the absorption power may be more easily accessible: 
With increasing $\phi$, the absorption maximum is shifted from $\delta\tilde{\epsilon}$ 
to ${\rm min}(T_K, \delta\epsilon)$, while at the same time the absorption intensity 
strongly increases (see Fig.~\ref{fig:Kondo}).

\begin{figure}
\vspace{0.5cm}
\begin{center}
\includegraphics[width=.6\textwidth]{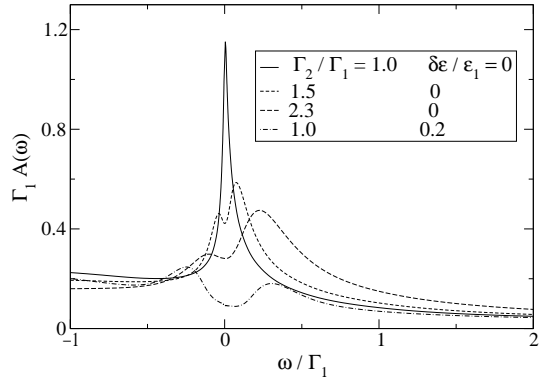}
\end{center}
\caption[]{Effect of an asymmetric broadening of the two levels 
for $U=16.2\Gamma_1$, $\epsilon_1 = -3.2\Gamma_1$, $\Phi = \pi$, $T=0$ 
and different values of $\Gamma_2$. 
For comparison, 
the influence of a finite level splitting is also shown.}
\label{fig:asymmetric}
\end{figure}

So far, we have assumed that both levels are equally broadened by the reservoir, 
i.e. $\Gamma_1 = \Gamma_2$. In experiments, this need not be the case, although tuning within $\delta\Gamma = 20\%$ seems feasible. In order to clarify whether the \emph{orbital} Kondo effect discussed 
previously is still visible under these conditions, we have therefore determined 
the influence of an asymmetric broadening on the single particle spectrum (Fig.~\ref{fig:asymmetric}). 
In the Kondo regime $\phi\approx\pi$ our model then corresponds to an effective 
Anderson Hamiltonian with spin--dependent hybridization, which is interesting 
in itself and has not been studied before.  
For the parameters used, we find a robust Kondo peak, which 
is split, but remains clearly visible 
even at an asymmetry of $\delta\Gamma = 100\%$. 
We therefore conclude that the orbital Kondo effect discussed here should be 
accessible under realistic experimental conditions. 

Finally, we would like to extend our model to include the electron spin, 
which should give rise also to a \emph{magnetic} Kondo effect.
In this case, the dot Hamiltonian has to be generalized as 
\beq
H_{\rm dot} = \sum_j \varepsilon_j c^\dagger_{j\sigma} c^\pdag_{j\sigma} + U n_1 n_2 + 
U_1 n_{1\uparrow} n_{1\downarrow} + U_2 n_{2\uparrow} n_{2\downarrow} 
\label{dot_with_spin}
\eeq
where $n_i = \sum_\sigma n_{i\sigma}$. 
For the special case of $U=U_1=U_2$ and $\delta\epsilon = 0$, this model has been studied 
before \cite{Izumida97,Izumida98}. In realistic double--dot systems, however, we 
expect $U < U_{1,2}$. This is the parameter regime we will address here. 

\begin{figure}
\begin{center}
\includegraphics[width=.5\textwidth]{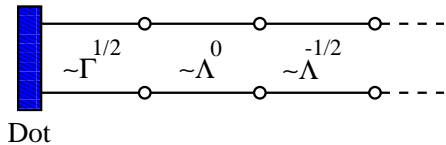}
\end{center}
\caption[]{Effective NRG Hamiltonian for the spin--degenerate dot (\ref{dot_with_spin}).}
\label{fig:double_chain}
\end{figure}

NRG calculations for the model including spin are very expensive from 
a computational point of view. This is due to the larger size of the conduction band 
Hilbert space, which now contains four instead of two 
additional fermionic degrees of freedom per iteration, as 
shown in Fig.~\ref{fig:double_chain}.
In our calculations, we kept about up to 1000 levels in each iteration step, which for 
20 iterations required about 10 hours of CPU time on an IBM Power 3.

\begin{figure}
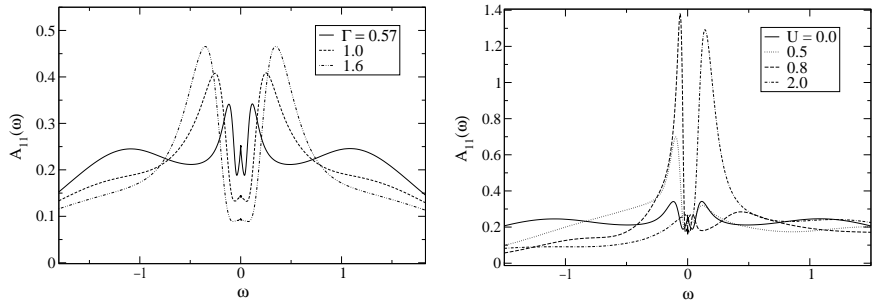

\begin{minipage}{0.5\linewidth}
\begin{center}
\includegraphics[width=.95\textwidth]{autfig14.eps}
\end{center}
\end{minipage}
\begin{minipage}{0.5\linewidth}
\begin{center}
\includegraphics[width=.95\textwidth]{autfig15.eps}
\end{center}
\end{minipage}
\caption[]{Partial spectral density for level $c_1$ of a spin--degenerate 
double quantum dot. 
Left: $U=0$, $U_1 = U_2 = 2.0$, $\epsilon_1 = \epsilon_2 = -1.0$ and 
$\Phi = T = 0$ with varying $\Gamma$.
Right: Same parameters, but now $\Gamma = 0.61$ is fixed and the interdot 
interaction $U$ is tuned to different values.}
\label{fig:kondo+interferenz}
\end{figure}

Results are shown in Fig.~\ref{fig:kondo+interferenz}: 
For vanishing AB--phase $\phi$, the two dot levels are coupled to the same 
reservoir. The resulting RKKY interaction \cite{Ruderman54} leads to an effective ferromagnetic 
exchange coupling between the levels.
For zero interdot correlation ($U=0$), the resulting side peaks dominate 
the single--particle spectrum (left plot). In addition, the screening of 
the total dot spin leads to the Kondo resonance at the Fermi level, which 
is suppressed for increasing broadening $\Gamma$. 

Switching on the interdot $U$ as shown in the right plot of 
Fig.~\ref{fig:kondo+interferenz} enhances the lower side peak. 
At $U=1$, a discontinuous change in the ground state occupation number occurs, 
while at even larger values of $U$, one again obtains the interference ``shoulder'' already 
discussed in the spinless case. 

For a finite AB--phase $\phi$ we expect interplay of the \emph{orbital} 
and \emph{magnetic} Kondo effects, which will be the subject of a future 
publication \cite{Hofstetter01}.

\section{Conclusion}
In this paper we have discussed Wilson's Numerical Renormalization Group, 
a powerful nonperturbative method designed specifically to 
calculate low--temperature properties of quantum impurity systems. 
It is the only method that yields reliable results for 
systems with very different energy scales (small Kondo temperature $T_K$, 
large bandwidth). 
Recent experimental progress in the fabrication of ultrasmall quantum dots 
has made the preparation of artificial Kondo ``atoms'' with well--controlled 
parameters possible. 
NRG is the method of choice for the theoretical interpretation 
of spectral and transport measurements in terms of single impurity models. 

We have presented an extended NRG algorithm (DM--NRG) 
suitable for calculating low--temperature dynamics in the full frequency range. 
This method has been applied to calculate the spectrum of a quantum dot in a 
magnetic field. 

Furthermore, we have studied the interplay between interference and Kondo correlations 
in multi--level quantum dots. \emph{Orbital} Kondo screening has been 
observed which can be tuned by an external Aharonov--Bohm phase and should be 
most easily visible in the optical absorption power. 
Additional spin degeneracy gives rise to the (Spin--) Kondo effect and 
causes an effective RKKY--interaction between the two dot levels. 
We expect competition between magnetic and orbital screening in the 
presence of a finite AB--phase. 
These results should be useful for the interpretation of recent experiments 
on vertically coupled dots (see e.g.~\cite{Wilhelm00}) or lateral 
multi--dot arrangements \cite{Holleitner00}. 

In future applications of the NRG method, more complex impurities
will be considered -- one may even try to model the lowest exitations 
of a complete molecule in order to describe recent transport experiments, see 
e.g.~\cite{Kergueris99}. 
From the methodical point of view, the extension of NRG to non--equilibrium 
calculations remains a major challenge. 

The author would like to thank R.~Bulla, H.~Schoeller, and D.~Vollhardt 
for valuable discussions. This work was supported by the 
Deutsche Forschungsgemeinschaft through SFB 484.

%INDEX%%%%%%%%%%%%%%%%%%%%%%%%%%%%%%%%%%%%%%%%%%%%%%%%%%%%%%%%%%%%%%%
% Please code your entries to include a "mutual" subject index in the
% standard syntax. For your own purposes you may print your
% "personal" index by using the following commands:
%
%\clearpage
%\addcontentsline{toc}{section}{Index}
%\flushbottom
%\printindex
%%%%%%%%%%%%%%%%%%%%%%%%%%%%%%%%%%%%%%%%%%%%%%%%%%%%%%%%%%%%%%%%%%%%%


\begin{thebibliography}{8.}
\addcontentsline{toc}{section}{References}

\bibitem{Hewson93}
A.C.~Hewson, {\em The Kondo Problem to Heavy Fermions}, 
(Cambridge University Press, 1993).

\bibitem{kondo-exp}
D.~Goldhaber-Gordon {\em et al.}, Nature {\bf 391}, 156 (1998);
S.M.~Cronenwett {\em et al.}, Science {\bf 281}, 540 (1998);
J.~Schmid {\em et al.}, Phys. Rev. Lett. {\bf 84}, 5824 (2000).
W.~G.~van der Wiel {\em et al.}, Science {\bf 289}, 2105 (2000).

\bibitem{Li98}
J.~Li, W.-D.~Schneider, R.~Berndt, and B.~Delley, Phys.~Rev.~Lett.~{\bf 80}, 2893 (1998).

\bibitem{Manoharan00}
H.C.~Manoharan, C.P.~Lutz, and D.M.~Eigler, Nature {\bf 403}, 512 (2000).

\bibitem{Anderson61}
P.W.~Anderson, Phys.~Rev.~{\bf B 124}, 41 (1961).

\bibitem{Kondo64}
J.~Kondo, Prog.~Theor.~Phys.~{\bf 32}, 37 (1964).

\bibitem{Metzner89}
W.~Metzner and D.~Vollhardt, Phys. Rev. Lett. {\bf 62}, 324 (1989).

\bibitem{Georges96}
A.~Georges, G.~Kotliar, W.~Krauth, and M.J.~Rozenberg, Rev. Mod. Phys. {\bf 68}, 13 (1996).

\bibitem{curacao}
{\em Mesoscopic Electron Transport}, eds. L.L.~Sohn {\em et al.}, 
(Kluwer, 1997); {\em Single Charge Tunneling}, eds. H.~Grabert
and M.H.~Devoret, (Plenum Press, 1991); D.V.~Averin and K.K.~Likharev,
in {\em Mesoscopic Phenomena in Solids}, ed. B.L.~Altshuler,
(Elsevier, 1991).

\bibitem{Simmel99}
F.~Simmel {\em et al.}, Phys. Rev. Lett. {\bf 83}, 804 (1999).

\bibitem{kondo-theo}
L.I.~Glazman and M.E.~Raikh, Sov. Phys. JETP Lett. {\bf 47}, 452 (1988); 
T.K.~Ng, P.A.~Lee, Phys. Rev. Lett. {\bf 61}, 1768 (1988); Y.~Meir,
N.S.~Wingreen, and P.A.~Lee, Phys. Rev. Lett. {\bf 70}, 2601 (1993).

\bibitem{Wilson75}
K.~G.~Wilson, Rev. Mod. Phys. {\bf 47}, 773 (1975).

\bibitem{Krishnamurthy80} 
H.R.~Krishna-Murthy, J.W.~Wilkins and K.G.~Wilson, Phys. Rev. {\bf B 21}, 1003 (1980).

\bibitem{Sakai89}
O.~Sakai, Y.~Shimizu, and T.~Kasuya, J.~Phys.~Soc.~Jpn.~{\bf 58}, 3666 (1989).

\bibitem{Costi94}
T.A.~Costi, A.C. Hewson, and V. Zlati\'{c}, J. Phys.:Cond. Mat. {\bf 6}, 2519 (1994).

\bibitem{Bulla98}
\mbox{R.~Bulla}, \mbox{A.C.~Hewson}, and \mbox{Th.~Pruschke}, 
J.~Phys.: Cond.~Mat.~{\bf 10}, 8365 (1998).

\bibitem{Hofstetter00b}
W.~Hofstetter, R.~Bulla, and D.~Vollhardt, Phys. Rev. Lett. {\bf 84}, 4417 (2000).

\bibitem{Hofstetter00a}
W.~Hofstetter, Phys. Rev. Lett. {\bf 85}, 1508 (2000).

\bibitem{Hofstetter00c}
Walter Hofstetter, \emph{Renormalization Group Methods for Quantum Impurity Systems}, 
(Shaker, Aachen 2000).

\bibitem{Feynman72}
R.P.~Feynman, \emph{Statistical Mechanics}, Addison Wesley, New York 1972.

\bibitem{Costi00}
T.A.~Costi, Phys. Rev. Lett. 85, 1504 (2000).

\bibitem{Yacoby97}
A.~Yacoby {\em et al.}, Phys. Rev. Lett. {\bf 74}, 4047 (1995);
R.~Schuster {\em et al.}, Nature {\bf 385}, 417 (1997).

\bibitem{loss}
D.~Loss and E.V.~Sukhorukov, Phys. Rev. Lett. {\bf 84}, 1035 (2000).

\bibitem{Boese00}
D.~Boese, W.~Hofstetter, and H.~Schoeller, preprint cond-mat/0010250.

\bibitem{Inoshita93}
T.~Inoshita {\em et al.}, Phys. Rev. B {\bf 48}, 14725 (1993).

\bibitem{Pohjola97}
T.~Pohjola {\em et al.}, Europhys. Lett. {\bf 40}, 189-194 (1997).

\bibitem{Izumida97}
W.~Izumida {\em et al.}, J. Phys. Soc. Jpn. {\bf 66}, 717 (1997).

\bibitem{Izumida98}
W.~Izumida {\em et al.}, J. Phys. Soc. Jpn. {\bf 67}, 2444 (1998).

\bibitem{Ruderman54}M.A.~Ruderman and C.~Kittel, Phys. Rev. {\bf 96}, 
99 (1954).

\bibitem{Hofstetter01}
W.~Hofstetter, to be published.

\bibitem{Wilhelm00}
U.~Wilhelm and J.~Weis, Physica {\bf E6}, 668 (2000).

\bibitem{Holleitner00}
A.W. Holleitner \emph{et al.}, preprint cond-mat/0011044.

\bibitem{Kergueris99}
C.~Kergueris, J.-P.~Bourgoin, S.~Palacin, D.~Esteve, C.~Urbina, 
M.~Magoga, and C.~Joachim, Phys.~Rev.~{\bf B 59}, 12505 (1999).



\end{thebibliography}
\end{document}